# Hypothalamic expression analysis of m$^6$A RNA methylation associated genes suggests a potential role of epitransciptomics in sexual maturation of Atlantic salmon

**Authors:**

Ehsan Pashay Ahi [1*]
Email: ehsan.pashayahi@helsinki.fi

Morgane Frapin [1*]
Email: morgane.frapin@helsinki.fi

Mikaela Hukkanen [1,2]
Email: mikaela.hukkanen@helsinki.fi

Craig R. Primmer [1,3]
Email: craig.primmer@helsinki.fi

**Corresponding Author:**

Ehsan Pashay Ahi,
Email: ehsan.pashayahi@helsinki.fi

**\* These authors contributed equally to this study**

1. Organismal and Evolutionary Biology Research Programme, Faculty of Biological and Environmental Sciences, University of Helsinki, Viikinkaari 9, 00014, Helsinki, Finland

2. Institute for Molecular Medicine Finland (FIMM), Helsinki Institute of Life Science (HiLIFE), University of Helsinki, Helsinki, Finland

3. Institute of Biotechnology, Helsinki Institute of Life Science (HiLIFE), University of Helsinki, Finland

# Abstract

Better understanding the molecular processes contributing to variation in maturation timing is important for Atlantic salmon (*Salmo salar*) aquaculture, as early maturation causes considerable financial losses. The m$^6$A RNA methylation is a highly conserved and dynamically reversible mechanism controlling gene expression in a myriad of biological processes. The role of m$^6$A methylation in sexual maturation, however, has remained largely unexplored and has never been studied in Atlantic salmon. While the maturation process is known to be affected by many genetic and environmental factors, the molecular mechanisms causing variation in the timing of maturation are still poorly understood. Hence, investigation of whether a widespread mechanism like m$^6$A methylation could be involved in controlling of the maturation timing is warranted. In Atlantic salmon, two genes, also associated with age at maturity in humans, *vgll3* and *six6*, have been shown to play an important role in maturation timing. In this study, we investigated the expression of 16 genes involved in the regulation of m$^6$A RNA methylation in the hypothalamus of Atlantic salmon with different homozygous combinations of *late* (L) and *early* (E) alleles for *vgll3* and *six6* genes. We found differential expression of *ythdf2.2*, an m$^6$A reader promoting mRNA degradation, with higher expression in *six6\*LL* compared to other genotypes as well as in immature compared to mature males. In addition, we found that the expression levels of genes coding for an eraser, *alkbh5*, and for a reader, *ythdf1*, were higher in the hypothalamus of females than in males across all the different genotypes studied. However the total m$^6$A levels between the whole hypothalamus of males and females were similar. Our results indicate a potential role of the m$^6$A methylation process in sexual maturation of Atlantic salmon, and therefore, provide the first evidence for such regulatory mechanism in the hypothalamus of any vertebrate.

# Introduction

In Atlantic salmon aquaculture, better understanding of the molecular processes contributing to variation in maturation timing is highly important, as early maturation in the form of grilsing (maturation one year after smoltification) causes considerable financial losses (McClure et al., 2007; Taranger et al., 2010) due to flesh quality degradation and decreased growth, and also has negative impacts on fish welfare (Aksnes et al., 1986; Good and Davidson, 2016; Taranger et al., 2010). Interestingly, two genes that have been associated with age at maturity in human, *VGLL3* (Vestigial Like Family Member 3) and *SIX6* (*sine oculis* homeobox 6) (Cousminer et al., 2016; Perry et al., 2014), also play a major role in pubertal timing of Atlantic salmon (*vgll3* and *six6*) (Barson et al., 2015; Sinclair-Waters et al., 2020). The *vgll3* genotype is strongly associated with maturation timing in both sexes of wild Atlantic salmon but also exhibits sex-specific maturation effects (Barson et al., 2015; Czorlich et al., 2018). This association between *vgll3* genotype and maturation probability has been validated in one year-old male parr in common garden settings (Debes et al., 2021; Sinclair-Waters et al., 2021; Verta et al., 2020). In our recent studies, we also showed that *vgll3* strongly affects the expression of reproductive axis genes in one year-old male Atlantic salmon (Ahi et al., 2022), and its regulatory effects on transcription of the gonadotropin encoding genes (*fshb* and *lhb*) are predicted to be mediated by the Hippo pathway (Ahi et al., 2023); a molecular pathway involved in various biological processes including a role in sexual maturation in both sexes (Clark et al., 2022; Lalonde-Larue et al., 2022). The molecular mechanism by which *six6* regulates pubertal timing is not known, but it seems that *six6* transcriptional regulation is not linked to *vgll3* function and it is independent of Hippo signaling (Kurko et al., 2020). The molecular mechanisms behind the associations of *vgll3* and *six6* genotypes with the age at maturity in Atlantic salmon remain therefore largely under-explored.

The epigenetic processes with potential roles in regulation of maturation timing are among prime candidate mechanisms to investigate, since (I) they affect numerous major biological processes during the ontogeny of an organism, (II) they have remained poorly studied in respect to regulation of sexual maturation, (III) they can provide molecular links between genes with seemingly unrelated functions, and (IV) they offer potential for breakthrough applications in aquaculture (e.g. in disease control and enhancing economically important traits) due to their inducible and reversible nature in regulation of gene expression without altering the genome (Roy et al., 2021). With respect to aquaculture, for instance, parental conditioning through RNA methylation has been shown to create transgenerational innate immune memory in shrimp without a need for genetic interference (Roy et al., 2022). One of these epigenetic mechanisms is the methylation at the N6 position of adenosine of RNAs (N6-Methyladenosine, hereafter m$^6$A) (Shi et al., 2022). Being the most abundant, reversible and highly conserved mRNA modification in eukaryotes, m$^6$A has recently emerged as a universal regulatory mechanism controlling gene expression in myriad of biological processes (Zaccara et al., 2019). The N6-Methyladenosine, m$^6$A, controls the fate of mRNAs within cells by acting on processes such as mRNA stability, splicing and transport. These modifications are added by the m$^6$A methyltransferase complex, which includes proteins called writers (e.g. Mettl3, Mettl14 and Wtap), and can be removed by demethylases called erasers (e.g. Fto and Alkbh5) (Zaccara et al., 2019). The RNA m$^6$A modifications are recognized by proteins called readers (e.g. Ythdf1/2/3 and Ythdc1/2) that guide methylated mRNA towards specific fates such as degradation, stabilization, transportation, and promotion or inhibition of translation (Liao et al., 2018).

Investigating the role of m$^6$A RNA modifications on economically important phenotypes, such as maturation timing, in commercial species like Atlantic salmon, could lead to useful applications in aquaculture. The m$^6$A RNA modification has been recently found to be crucially involved in reproduction, and findings on the essential role of m$^6$A modification during gametogenesis (Lasman et al., 2020; Li et al., 2021; Xia et al., 2018) may be just the tip of the iceberg. However, currently there is no knowledge on the potential role of m$^6$A modification upstream of initiating and orchestrating sexual maturation along hypothalamus-pituitary-gonadal (HPG) axis, and hence its possible role in contributing to variation in maturation timing. Furthermore, earlier pioneering studies on the role of the m$^6$A RNA modification in fish sexual maturation have focused on model species in laboratory settings and thus the significance of these mechanisms in commercially important salmonids remains to be confirmed (Roy et al., 2021). The complex molecular mechanisms causing variation in maturation timing are generally poorly understood (Howard and Dunkel, 2019; Leka-Emiri et al., 2017; Mobley et al., 2021), which makes it even more tempting to investigate whether a universal mechanism such as m$^6$A RNA modification may contribute to maturation timing variation.

In this study, we investigate the expression of 16 genes that have been found to code proteins associated with the regulation of m$^6$A modifications including 3 writers; *mettl3*, *mettl14*, and *wtap*, 4 erasers; *alkbh5-1*, *alkbh5-2*, *fto-1* and *fto-2*, and 9 readers; *ythdc1-1*, *ythdc1-2*, *ythdc2*, *ythdf1-1*,

*ythdf1-2*, *ythdf1-3*, *ythdf2-1*, *ythdf2-2* and *ythdf3*. We compare their mRNA expression between different homozygous combinations of *late* (L) and *early* (E) alleles for *vgll3* and *six6* genes in the hypothalamus between and within sexes of one year old Atlantic salmon. In males, we also compared the expression of m$^6$A methylation markers between immature and mature individuals. In order to elucidate the possible molecular mechanisms (regulatory role) of the studied genes in contributing to the timing of sexual maturation even further, we also studied directly the total levels of m$^6$A in both sexes. To our knowledge, this is the first study addressing differences in the expression of m$^6$A methylation markers in respect to sexual maturation in the hypothalamus.

## Materials and methods

### Fish rearing, genotyping and tissue sampling

The Atlantic salmon used in this study were created using parental individuals from a 1$^{st}$ generation hatchery broodstock originating from the Iijoki population that is maintained at the Natural Resources Institute (Luke) Taivaloski hatchery in northern Finland. Unrelated parents were chosen from broodstock individuals that had earlier been genotyped for 177 SNPs on Ion Torrent or Illumina (Miseq or Next-Seq) sequencing platforms (Aykanat et al., 2016). These SNPs included those associated with the unlinked *vgll3* and *six6* genes that were earlier shown to be linked to age at maturity in salmon (Barson et al., 2015). We selected parents that were heterozygotes for both *vgll3* and *six6* (*vgll3\*EL* and *six6\*EL*) in order for full-sib families to contain offspring with all *vgll3* - *six6* genotype combinations. We avoided crossing closely related individuals (those with grandparents in common) by using SNP-based pedigree reconstruction (Debes et al., 2020). In order to minimize unwanted variation, all individuals in this study originate from a single full-sib family. For simplicity, only the four alternative two locus homozygous genotype combinations (*vgll3\*EE* or *vgll3\*LL* and *six6\*EE* or *six6\*LL*) were examined. This enabled the assessment of the expression patterns of all the possible homozygous *vgll3* and *six6* genotypes within an otherwise similar genetic background. Other rearing, tagging and genotyping details are as described in Sinclair-Waters et al. (2021). From this point onwards, four character genotypes will be used to describe an individual's genotype at the focal loci, *vgll3* and *six6*. The first two characters indicate the genotype at the *vgll3* locus and the last two characters indicate the genotype at the *six6* locus. The locus is indicated in subscript text after the genotype.

The fish were euthanized approximately one-year post-fertilization during the spawning season in November with an overdose of the anesthetic buffered tricaine methane sulfonate (MS-222) and dissected, and sex and maturation status were determined visually by observing the presence of female or male gonads as outlined in Verta et al. (2020). The maturation status of males was classified as immature (no phenotypic signs of gonad maturation) and mature (large gonads leaking milt). The mature or immature status of male salmon was determined by respectively the presence or the absence of sperm leakage. At this age all the females were immature. Whole hypothalami of salmon with the genotypes of interest were dissected and snap-frozen in liquid nitrogen before being stored at -80 °C.

**RNA extraction and cDNA synthesis**

RNA was isolated using NucleoSpin RNA kit (Macherey-Nagel GmbH & Co. KG). The hypothalami were transferred to tubes with 1.4 mm ceramic beads (Omni International), Buffer RA1 and DDT (350ul RA1 and 3,5ul DDT 1M) and homogenized using Bead Ruptor Elite (Omni International) with tissue specific program (4m/s, 3x20s). Remaining steps of RNA isolation were conducted as in the manufacturer's protocol. RNA was eluted in 40 µl of nuclease free water. Quality and concentration of the samples were measured with NanoDrop ND-1000 and RNA integrity was assessed by agarose gel electrophoresis. The extracted RNA (400ng for females, 500ng for males per sample) was subsequently reverse-transcribed to cDNA using iScript cDNA Synthesis Kit (Bio-Rad Laboratories, Inc.).

**Primer design and qPCR**

A Whole Genome Duplication (WGD) that occurred ~80 million years ago in salmonids leads to the presence of paralogous genes (Lien et al., 2016). We used paralogue-specific gene sequences obtained from the recently annotated *Salmo salar* genome in the Ensembl database, http://www.ensembl.org. The paralogue gene sequences were aligned using CLC Genomic

Workbench (CLC Bio, Denmark) in order to identify paralogue specific regions for designing the primers. The paralogue specific primers were designed in the 3' end which is known to evolve more rapidly, thus showing the greatest variation between the paralogues (Bricout et al., 2023). In addition, the primers were designed to have different amplicon size for each paralogue, and consequently their melting curves would be different. Therefore, by conducting a melting curve analysis at the end of the amplification cycles of each qPCR, we were able to confirm that there was only one melting curve for each primer pairs. The steps of primer design, e.g., melting temperature and dimerization check, exon/exon junction spanning and short amplicon size, were performed as described in Ahi and Sefc, 2018, using two online tools: Primer Express 3.0 (Applied Biosystems, CA, USA) and OligoAnalyzer 3.1 (Integrated DNA Technology) (Supplementary Data). The qPCR reactions were prepared with two technical replicates in then following program of 2 min/50 °C + 10 min/ 95 °C; 40 x 15 sec/95 °C and 1 min/60 °C (as described in Ahi et al., 2019), using PowerUp SYBR Green Master Mix (Thermo Fischer Scientific), and performed on a Bio-Rad CFX96 Touch Real Time PCR Detection system (Bio-Rad, Hercules, CA, USA). The details of the qPCR program and calculation of primer efficiencies are described in Ahi et al., 2019.

In qPCR-based gene expression analyses, comparisons with stably expressed reference genes (Kubista et al., 2006) are necessary. These reference genes have to be validated for the specific species, tissue, and experimental conditions in each study (Ahi et al., 2013). We implemented three common algorithms to validate the most suitable reference gene(s); BestKeeper (Pfaffl et al., 2004), NormFinder (Andersen et al., 2004) and geNorm (Vandesompele et al., 2002). These algorithms use different analysis approaches to rank the most stably expressed reference genes. The Cq values of the target genes were normalized with the geometric mean of the Cq values of two top ranked references genes, *ef1a* and *hprt1*, following the formula $\Delta Cq_{target} = Cq_{target} - Cq_{reference}$. These $\Delta Cq_{target}$ values have been adjusted for qPCR batch effect using ComBat (v3) (Johnson et al., 2007) from sva R package (Surrogate Variable Analysis, v3.40.0) (Leek et al., 2012). The following calculations have been made on the qPCR batch adjusted values. For each gene, a biological replicate with the lowest expression level across all the samples used in each comparison (calibrator sample) was selected to calculate $\Delta\Delta Cq$ values ($\Delta Cq_{target} - \Delta Cq_{calibrator}$). The relative expression quantities (RQ values) were calculated as $2^{-\Delta\Delta Cq}$, and their fold changes (logarithmic values of RQs) were used for statistical analysis (Pfaffl, 2001).

**Quantification of total m⁶A RNA methylation**

The total m$^6$A levels in RNA were measured with the EpiQuick™ m$^6$A RNA methylation Quantification kit (Colorimetric) (EpiGentek) according to the manufacturer's protocol. The samples as well as positive and negative controls were analyzed in duplicate. The average OD450 value of the duplicates was used to calculate the percentage of m$^6$A in total RNA and it was calculated based on the relative quantification calculation in the manufacturer's protocol. The amount of input RNA was determined based on concentrations measured with Qubit™ Broad Range assay kit (Invitrogen™) and was approximately 200 ng.

**Statistics**

For qPCR data, the effects of genotype and maturity status were modeled separately: the between-genotype effects were examined within maturity stages, whereas the between-maturity stages were examined within each genotype. An ANOVA (Analysis of variance) test was applied for the analysis between genotypes. Benjamini-Hochberg method (Benjamini and Hochberg, 1995) was used for multiple-comparison correction. In cases where the variable of interest (genotype or maturity) explained variation significantly (FDR-adjusted p-value < 0.05) after correcting for multiple comparisons, further post-hoc tests were conducted. Post-hoc tests were calculated using Tukey's Honest Significant Difference test.

For m$^6$A methylation quantification, to determine if there was a significant difference in the percentage of m$^6$A in total RNA between males and females, a Student's t-test was performed. The decision to use a parametric test was made after testing the normal distribution and homogeneity of variance of the data with the Shapiro-Wilk and Levene's tests, respectively. A significance threshold of 0.05 was used (i.e., an alpha level of 0.05), and p-values less than this threshold were considered significant. Statistical analyses were conducted using R software (version 4.1.1; R Core Team, 2023).

**Results**

**Validation of reference genes in in hypothalamus of Atlantic salmon**

We tested the expression of six reference gene candidates, which are commonly used in brain tissues of Atlantic salmon (Ahi et al., 2022) across all the hypothalamus samples in this study. The genes were ranked based on three algorithms, *i.e.* BestKeeper, geNorm and NormFinder (Table 1). Among the reference genes, *hprt1* ranked first by geNorm and NormFinder and second by BestKeeper analyses (Table 1). In addition, *ef1a* was ranked first by Bestkeeper and second by geNorm and NormFinder. Therefore, the data indicated high expression stabilities for of *hprt1* and *ef1a* and we used geometric mean of the expression for both genes as normalization factor to accurately quantify differences in gene expression between the samples.

**Expression levels of m$^6$A methylation markers in hypothalamus of Atlantic salmon**

We first assessed the overall expression levels of genes encoding for proteins involved in the regulation of m6A methylation in males (maturity status separately) and in females (Fig. 1). Among the writers, *wtap* had the highest expression level (lowest dCq values), whereas *mettl3* displayed the lowest expression level (highest dCq values) (Fig. 1A). The paralogous genes of the two erasers showed different expression levels as well, i.e. *alkbh5-1* > *alkhb5-2* and *fto-1* > *fto-2* (Fig. 1B). Similarly, the paralogous genes of three readers showed differences in their expression level, i.e. *ythdc1-2* > *ythdc1-1*, *ythdf1-3* > *ythdf1-2* > *ythdf1-1*, and *ythdf2-1* > *ythdf2-2* (Fig. 1C). However, the expression level difference between *ythdf2-1* and *ythdf2-2* appeared to be minor compared to the differences between the paralogs of the other genes. Among the readers, *ythdf1-3* had the highest expression level, whereas *ythdc2* exhibited the lowest expression level (Fig. 1C). In general, the sex and maturity status did not seem to have any effect on expression level differences between the paralogs.

**Expression differences of m$^6$A methylation markers between the genotypes**

Next, we explored the expression differences of the m6A methylation regulators between the genotypes within each sex and maturity status (in males). In immature males, we found expression differences between the genotypes for only one paralogue of a reader gene, *ythdf2-2*, which displayed higher expression in the genotype combinations with homozygous *late* allele of *six6* (*six6\*LL*) (Fig. 2). We did not observe any *vgll3* or *six6* genotype-specific expression difference in the hypothalamus of either mature males or immature females (Supplementary Fig. 1 and 2).

**Expression differences of m$^6$A methylation markers between the maturity statuses in males**

The comparison between maturity stages was only possible within two genotype combinations (*vgll3\*EE/six6\*LL* and *vgll3\*LL/six6\*EE*), as other genotype combinations did not have both immature and mature males due to the strong influence of the genes on maturation (all the individuals for *vgll3\*LL/six6\*LL* were immature, whereas for *vgll3\*EE/six6\*EE,* all were mature). Within *vgll3\*EE/six6\*LL* individuals, we observed higher expression of *ythdf2-2* in the hypothalami of immature males compared to mature males (Fig. 3). Whereas, within *vgll3\*LL/six6\*EE* genotype, we found a paralog of another reader gene, *ythdf1-2*, showing similar expression pattern with higher expression in the hypothalami of immature males compared to mature males (Fig. 4)**.**

**Differences in expression of m$^6$A methylation markers and amount of total m$^6$A between the sexes**

Finally, we compared the expression levels of the m$^6$A methylation regulators between sexes within the genotypes classes containing both immature males and females (Fig. 5). We found a paralogous gene of an eraser *alkbh5-1* and a paralogous gene of a reader, *ythdf1-3*, to be differentially expressed between the males and females in all the genotypes. Interestingly, the direction of these expression pattern differences was always the same for both genes; showing higher expression levels in the hypothalamus of females than males. Also, another paralogous gene of the same reader, *ythdf1-2*, showed similar tendency of higher expression in females but the differences were not statistically significant. These findings indicate sex-specific, but not genotype-specific, expression level differences for both of these genes in the hypothalamus of Atlantic salmon. Based on the higher RNA expression level of the eraser gene paralog, *alkbh5-1*, observed in immature females compared to immature males, we measured the percentage of m$^6$A in hypothalamic total RNA of 15 females and 18 males. The amount of m$^6$A was not significantly different between males and females (males: 0.092 ± 0.021 %, female: 0.094 ± 0.026 %, p-value = 0.811, t-test analysis).

**Discussion**

The expression of the m$^6$A RNA modification regulators in hypothalamus has never been the focus of research in fish reproduction to date. This is somewhat surprising, not only because the hypothalamus is an important upstream organ regulating sexual maturation (Kah et al., 1993), but also due to the widespread roles of m$^6$A RNA methylation, as the most abundant RNA modification in eukaryotic cells, in regulation of numerous biological processes (Zaccara et al., 2019). Particularly, m$^6$A RNA modification has been described as an essential mechanism in gonadal development and fertility (Mu et al., 2022). However, in the downstream sex organs, i.e. gonads, the role and expression of some of the m$^6$A RNA modification markers have been recently studied in fish (Wang et al., 2020; Xia et al., 2018; Zhao et al., 2021). Thus, in this study, we first set out to investigate, for the first time, whether the m$^6$A RNA modification regulators are expressed in hypothalamus of a commercially significant fish species — the Atlantic salmon. We found that at least 16 genes encoding the m$^6$A RNA modification regulators (writers, erasers and readers) are expressed in the hypothalamus but with varying levels. Importantly, it appeared that even the paralogs of the same eraser and reader genes can have very variable expression levels in the hypothalamus indicating potential differences in the functional importance of these genes at paralogue level.

The most important finding in this study was the differential expression of *ythdf2-2* in the hypothalamus of male Atlantic salmon, with reduced expression in mature compared to immature individuals with the *vgll3\*EE/six6\*LL* genotype combination, as well as reduced expression in *six6\*EE* genotype within immature groups. In Atlantic salmon, there are two paralogous genes (*ythdf2-1* and *ythdf2-2*) for *ythdf2*, encoding a member of YT521-B homology domain family (YTHDF) proteins, which acts as a reader of m$^6$A and its binding to m$^6$A-containing RNA leads to degradation of the RNA (Du et al., 2016). In mice, Ythdf2 deficiency results in female infertility and has pivotal role in maternal RNA degradation during oocyte maturation (Ivanova et al., 2017).

Another study in zebrafish also reported the pivotal role of *ythdf2* in the maternal-to-zygotic transition by orchestrating the clearance of almost one-third of maternal mRNAs in zygote (Zhao et al., 2017). More recent studies in mice have revealed that Ythdf2-mediated mRNA degradation on m$^6$A-modified target transcripts is required for spermatogenesis and fertility (Qi et al., 2022; Zhao et al., 2021). However, no study has investigated the potential role of *ythdf2* upstream of HPG axis (i.e. hypothalamus) during sexual maturation in both sexes. In mice, Ythdf2 has been shown to be essential during embryonic neural development by promoting m$^6$A-dependent degradation of genes related to neuron differentiation (Li et al., 2018). In chicken hypothalamus, RNA m$^6$A modification seems to be involved in the regulation of circadian rhythms under stressful conditions, and the expression m$^6$A methylation markers including *Ythdf2* show correlated oscillation with the clock genes (Y. Yang et al., 2022a). Studies in chickens and rats have shown that the level of expression of *Ythdf2* is modified by the environment (i.e. light exposure and maternal diet during gestation) in the hypothalamus (Frapin et al., 2020; Yang et al., 2022b). Interestingly, even though the direct hypothalamic role of *Ythdf2* during sexual maturation has not been explored yet, a decrease in the level of m$^6$A methylation by increased hypothalamic expression of *Fto* (a m6A eraser) has been shown to cause early onset of puberty in female rat (Yang et al., 2022). The opposite effect of delayed puberty was also obtained by knockdown of *Fto* (X. Yang et al., 2022), which, however, did not have differential hypothalamic expression with regards to sexual maturation in our study. The result of the latter study on female rat together with our findings about *ythdf2* indicate that m$^6$A modification might be involved in mediated sexual maturation in vertebrates by fine tuning the hypothalamic mRNA levels either through increased expression of a m$^6$A eraser (e.g. *Fto* in rat) or reduced expression of m$^6$A reader (*ythdf2* in Atlantic salmon).

The genotype dependent expression difference of *ythdf2* was also observed within immature individuals. This expression difference was more likely to be linked to *six6* genotype than *vgll3* genotype, since no difference was observed between *vgll3*LL* and *vgll3*EE* when *six6* genotype remained the same (i.e. it was fixed on *six6*LL*). In respect to *six6* genotype, however, *ythdf2* appeared to have higher expression in *six6*EE* than *six6*LL*. This can be functionally explained, as mentioned above, by the fact that puberty may favor reduced expression of *ythdf2* and increased level of its mRNA targets in hypothalamus and since *six6*EE* are more prone to enter puberty than *six6*LL*. But this does not explain why a similar pattern was not observed between *vgll3*LL* and

*vgll3\*EE*, unless there might be an unknown regulatory link only between *six6* and *ythdf2* and not between *vgll3* and *ythdf2*. In mammals, it has been shown that *six6* is involved in development of hypothalamic GnRH neurons, which are essential for the onset of puberty (Pandolfi et al., 2019). Moreover, Ythdf2-mediated mRNA clearance is also demonstrated to be important for neuron maturation in the developing forebrain of mice (Li et al., 2018). However, deeper understanding of the connection between *six6* genotypes and *ythdf2* expression requires further investigations.

Since no direct regulatory link was identified between *six6* and *ythdf2*, one plausible scenario is an indirect hierarchical regulatory connection between them. We have recently described such an indirect hierarchical regulatory connection in the pituitary of Atlantic salmon between *vgll3* and *jun*; an upstream regulator of sexual maturation (Ahi et al., 2023). In mammals, for instance, it is already known that *Six3* and *Six6* together induce the expression of the *Hes5* transcription repressor (Diacou et al., 2018), a negative regulator of neurogenesis which is highly expressed in the anterior part of developing hypothalamus (Aujla et al., 2015). A well-known direct downstream target of *Hes5* in brain is a gene called *Fbw7* or *Fbxw7* (Sancho et al., 2013), which encodes a member of the F-box protein family and controls neural stem cell differentiation in various parts of the brain (Hoeck et al., 2010). The transcriptional repression of *Fbw7* by Hes5 is essential for the correct specification of neural cell fates (Sancho et al., 2013). Interestingly, it has been recently shown that Ythdf2 is a direct substrate for Fbw7 and their interaction leads to proteolytic degradation of Ythdf2 (Xu et al., 2021). These findings in mammals indicate a potential hierarchical regulatory axis consisting of Six6/Hes5/Fbw7/Ythdf2 which can explain a molecular link between Six6 and Ythdf2. However, this model still does not explain the differential transcriptional regulation of *ythdf2* by distinct *six6* genotypes in the hypothalamus of Atlantic salmon.

Another interesting finding in this study was the sex-specific differential expression of *alkbh5-1* and *ythdf1-3* with higher expressions in the female hypothalamus. The protein encode by *alkbh5-1* is a well-known m$^6$A demethylase regulating RNA metabolism and nuclear export of mRNA and is involved in a variety of processes including spermatogenesis and male fertility (Tang et al., 2017; Zheng et al., 2013). The genes *ythdf1* encode for a m$^6$A reader which facilitates translation initiation (Wang et al., 2015). Although the specific role(s) of *Ythdf1* has not been explored in the hypothalamus, *Ythdf1* has been shown to play various roles in central nervous system such as hippocampus-dependent learning and memory processing (Shi et al., 2018), spinal axon guidance (Zhuang et al., 2019), cerebellar fiber growth (Yu et al., 2021), and nerve regeneration (Livneh et al., 2020). Interestingly, it is shown in mice that *Ythdf1* is a direct post-transcriptional target for Alkbh5 and its mRNA m$^6$A demethylation is mediated by eraser activity of Alkbh5 which leads to increased expression level of *Ythdf1* (Han et al., 2021). This is consistent with our findings, and the identical expression patterns between *alkbh5-1* and *ythdf1-3*, both with higher expression in the hypothalamus of females, suggests a regulatory connection between them in salmon similar to the one observed in mice.

However, in our study, the higher hypothalamic expression of *alkbh5-1* in female Atlantic salmon was not associated with an overall lower level of m$^6$A methylation, as might be expected from its role as a demethylase of m$^6$A. The hypothalamus is a tissue containing many different cell types and it has been suggested in mouse brain development that Alkbh5 might be primarily expressed in neurons (Du et al., 2020). The difference in m$^6$A levels between males and females could be cell type specific and therefore masked in the whole hypothalamus analysis. Another possibility which can't be detected by total m$^6$A quantification would be gene specific m$^6$A demethylase activity of *alkbh5-1* (Guo et al., 2020). The use of more in depth techniques such as MeRIP sequencing (previously called m$^6$A-seq (Meyer et al., 2012)) would be necessary to investigate such specific role.

The absence of an effect of a difference in *alkbh5-1* expression between immature males and females on the percentage of m$^6$A could also indicate regulatory divergence between the two paralogs which is a common event after whole genome duplication (Voordeckers et al., 2015). In this case, the role of the enzyme encoded by the *alkbh5-1* in the regulation of m$^6$A modifications may be different in Atlantic salmon compared to other species. Regulatory divergence of paralogs can also lead to a phenomenon called paralog interference, a common evolutionary constraining event after genome/gene duplication in which duplicated paralogs acquire competing/interfering functions (Baker et al., 2013). Paralog interference has been more often reported in genes encoding enzymes where slight structural changes in their products can result in opposing/neutralizing hetero-dimerization between them (Kaltenegger and Ober, 2015). Since *alkbh5* paralogs encodes enzymes, it is likely that they have experienced paralog interference, but this needs to be further investigated by structural and functional analyses.

## Conclusions

Though epigenetic reprogramming has huge economical potential, post-transcriptional epigenetic studies on important aquaculture species are still in their infancy. Understanding of epigenetic mechanisms involved in economically valuable phenotypic traits of aquaculture species will contribute to the selection of the most favourable ones. m$^6$A RNA modification may allow a further step forward by enabling possibilities to reversibly modify certain phenotypic traits without affecting the genetic sequence of aquaculture species. This study provides the first characterization of the level of m$^6$A methylation and expression patterns of various types of m$^6$A RNA modification regulator genes, i.e. writers, erasers and readers, in Atlantic salmon hypothalamus. Moreover, it demonstrates the existence of paralog-specific variation in expression of these genes in the hypothalamus of Atlantic salmon indicating potential differences in their function in this organ. Both genotype- and maturity status-dependent expression differences were detected for a well-known reader gene involved in mRNA degradation, *ythdf2-2*, i.e. higher expression in immature and *late* (L) puberty allele of *six6* in the hypothalamus of male Atlantic salmon. This suggests a potential role of *ythdf2-2* in sexual maturation in hypothalamus for the first time in a vertebrate species. In addition to *ythdf2-2*, we also found sex-specific expression (females > males) of an eraser gene, *alkbh5-1* and its downstream reader target, *ythdf1-3*, which may implicate an unknown sex-dependent mechanism regulating m$^6$A RNA modification regulator genes in the hypothalamus of the Atlantic salmon. Further transcriptional and functional investigations are required to understand the underlying regulatory mechanisms linking genotype, maturity status and sex to these m$^6$A RNA modification regulators.

## Acknowledgements

We acknowledge Jaakko Erkinaro and staff at the Natural Resources Institute Finland (Luke) hatchery in Laukaa and members of the Evolution, Conservation and Genomics research group for their help in coordinating and collecting gametes for crosses. We thank Nikolai Piavchenko, Claudius Kratochwil, Jacqueline Moustakas-Verho, Lea Ala-Ilomäki, Nea Asumaa, and Suvi Salonen for help with fish husbandry, and Valeria Valanne, Iikki Donner and Seija Tillanen for dissection and laboratory assistance.

## Author Contributions

EP, MF, CRP conceived the study; EP and MF performed experiments; MH, MF and EP developed methodology and analyzed the data; EP, MF and CRP interpreted results of the experiments; EP, MF and CRP drafted the manuscript, with EP having the main contribution, and all authors approved the final version of manuscript.

**Funding Source Declaration**

Funding was provided by Academy of Finland (grant numbers 307593, 302873, 327255 and 342851), and the European Research Council under the European Articles Union's Horizon 2020 research and innovation program (grant no. 742312).

**Competing financial interests**

Authors declare no competing interests

**Ethical approval**

Animal experimentation followed European Union Directive 2010/63/EU under license ESAVI/35841/2020 granted by the Animal Experiment Board in Finland (ELLA).

**Data availability**

All the gene expression data generated during this study are included in this article as supplementary file.

**Table 1. Expression stability ranking of reference genes in the hypothalamus of Atlantic salmon.** Abbreviations: NE = Not expressed, SV = stability value, M = mean value of stability.

| geNorm | | NormFinder | | BestKeeper | |
|---|---|---|---|---|---|
| Ranking | M values | Ranking | S values | Ranking | SD values |
| *hprt1* | 0.623 | *hprt1* | 0.151 | *ef1a* | 0.484 |
| *ef1a* | 0.689 | *ef1a* | 0.182 | *hprt1* | 0.509 |
| *actb1* | 0.719 | *actb1* | 0.243 | *actb1* | 0.640 |
| *rna18s* | 0.770 | *rna18s* | 0.306 | *g6pd* | 0.723 |
| *g6pd* | 0.879 | *g6pd* | 0.357 | *gapdh* | 0.798 |
| *gapdh* | 1.168 | *gapdh* | 0.484 | *rna18s* | 0.821 |

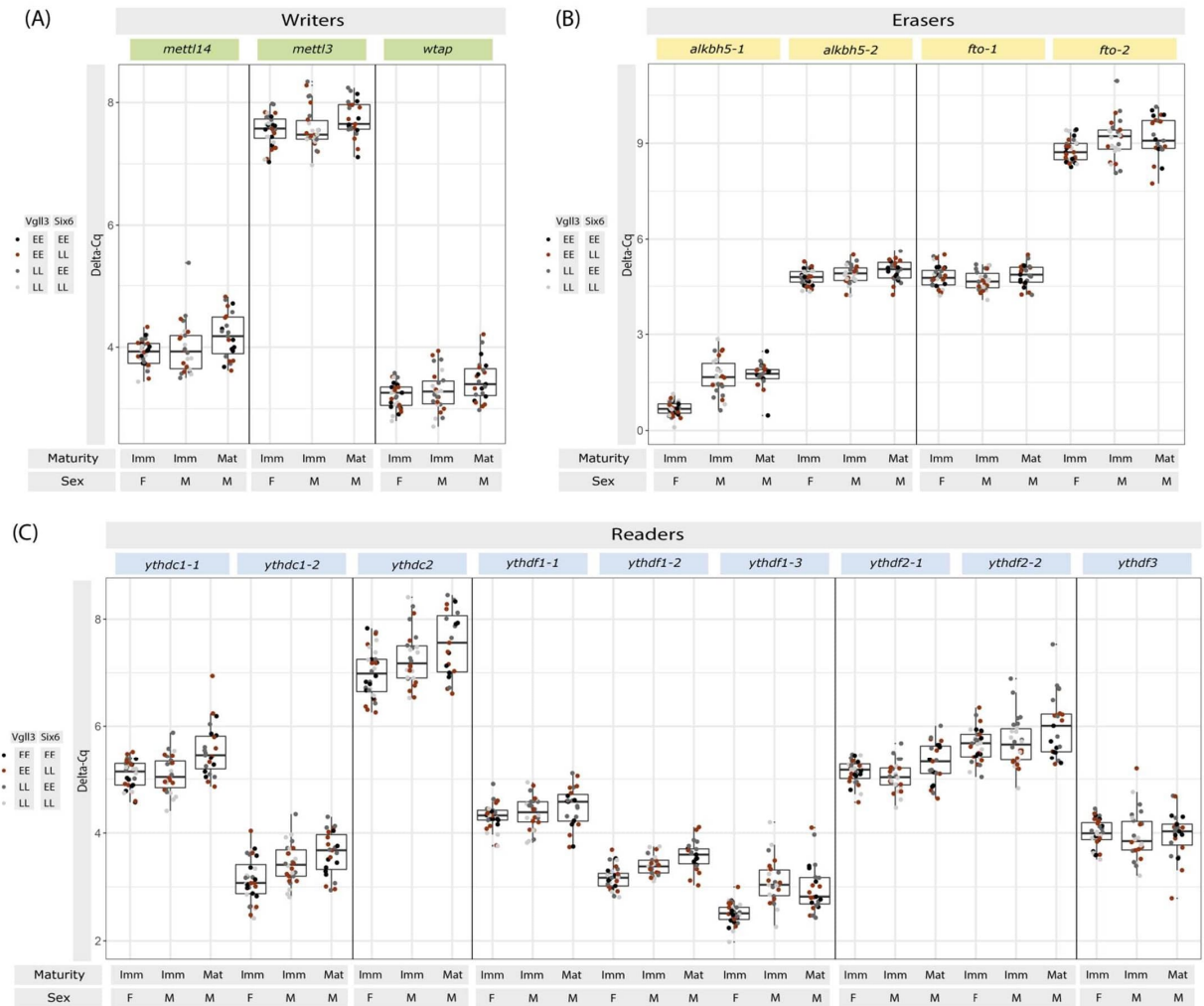

**Figure 1: Overall mRNA expression level of the m⁶A RNA modification regulators.** qPCR batch adjusted Delta-Cq of the genes coding for writers (A), erasers (B) and readers (C). The boxplots represent the median, first and third quartiles of all the genotypes within the group. (F: Females, M: Males, Imm: Immature, Mat: Mature).

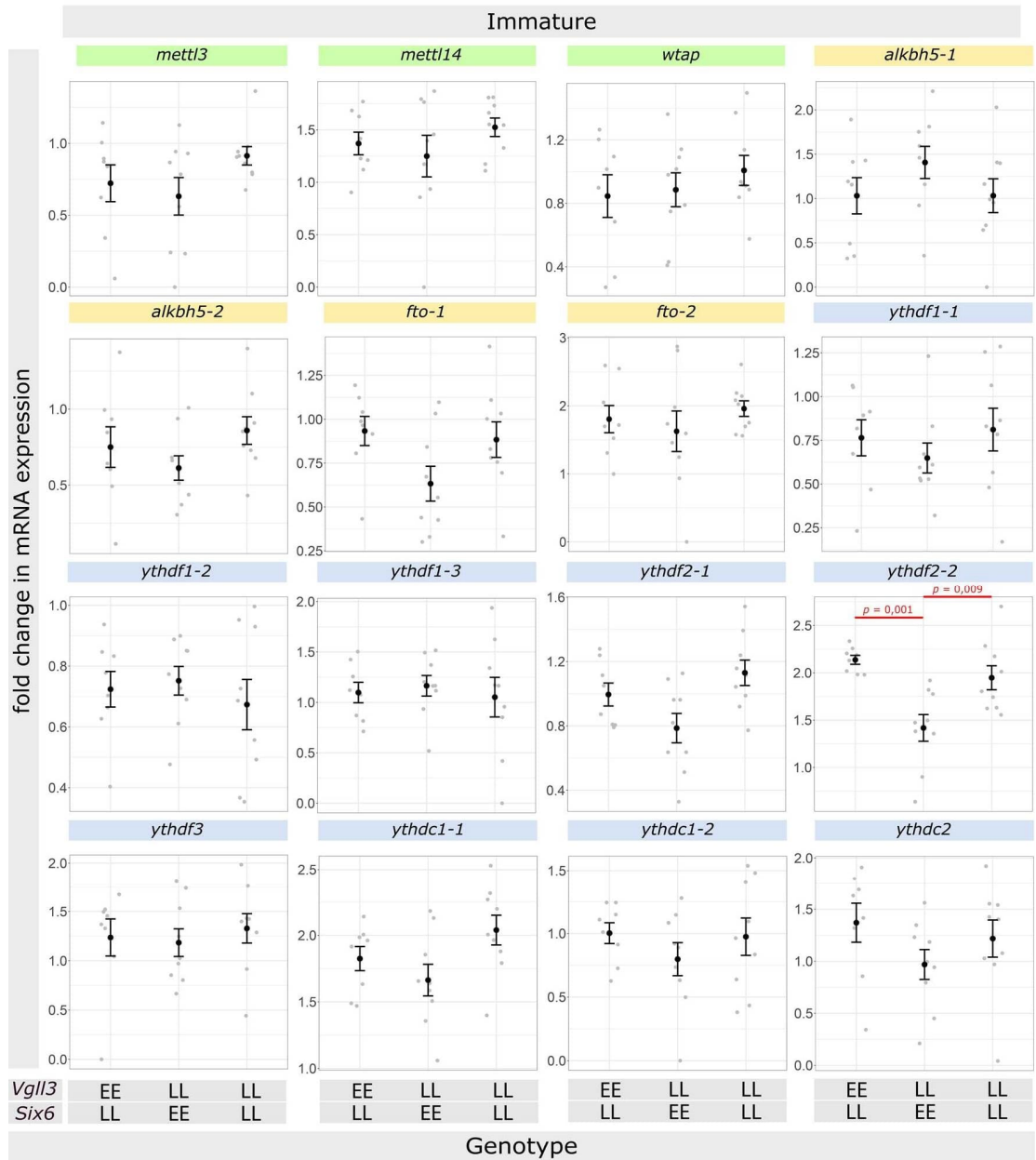

**Figure 2: Comparison of the m⁶A RNA modification regulators mRNA expression level between the genotypes in immature males.** For each gene, values were Log2 Fold-Change of the mRNA expression for each sample (grey dot) and mean ± SEM (black dot and bar). The mRNA level expression differences between the genotypes were analyzed with ANOVA and corrected for multiple comparison using the Benjamini-Hochberg method. p: p-values obtained from the Turkey HSD post-hoc test performed on the comparisons for which the FDR-adjusted p-values of the ANOVA test was < 0.05.

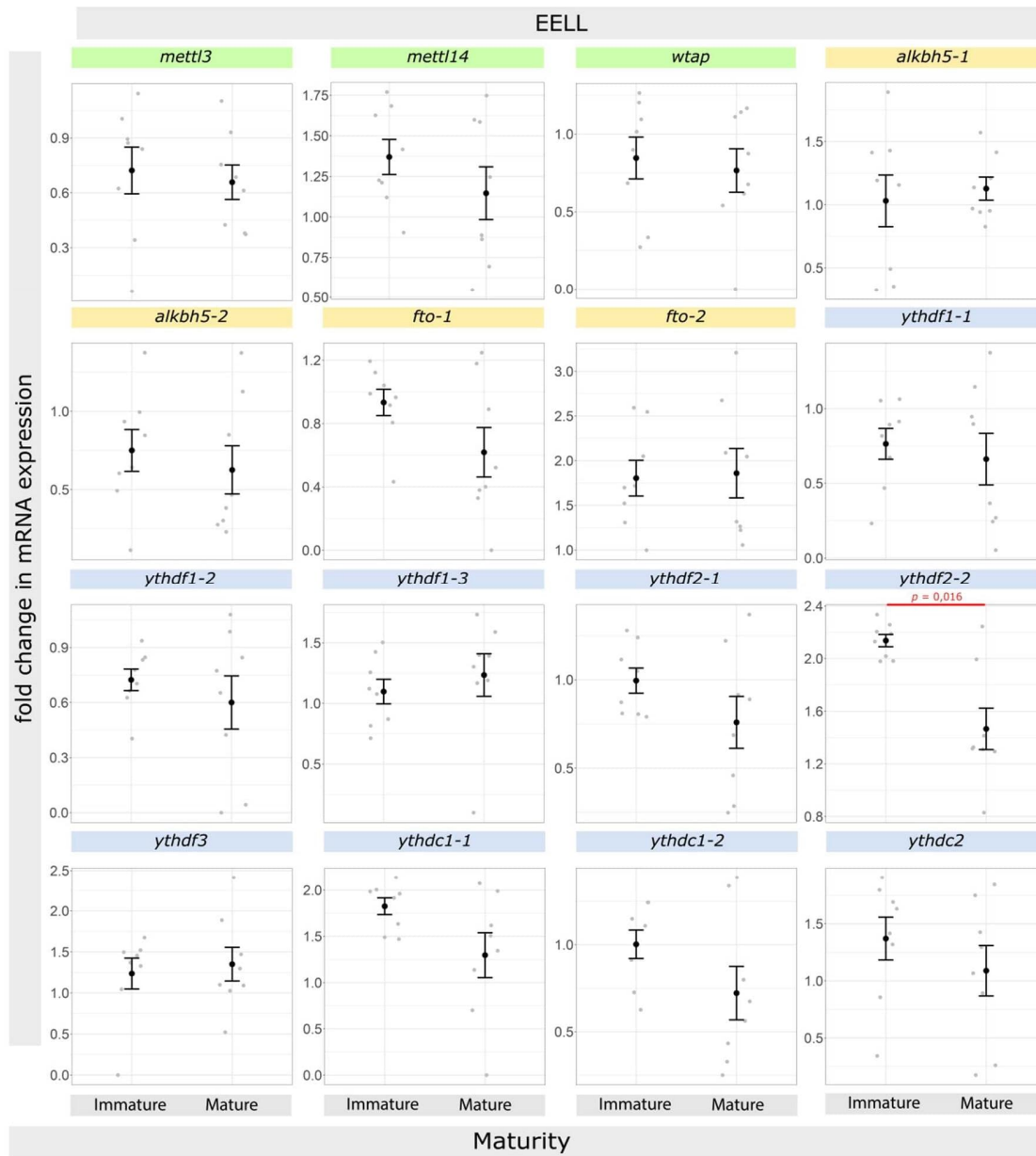

**Figure 3: Comparison of the m⁶A RNA modification regulators mRNA expression level between maturity stages in individuals with EE LL genotypes**. For each gene, values were Log2 Fold-Change of the mRNA expression for each sample (grey dot) and mean ± SEM (black dot and bar). The mRNA level expression differences between the genotypes were analyzed with ANOVA and corrected for multiple comparison using the Benjamini-Hochberg method. p: FDR-adjusted p-values.

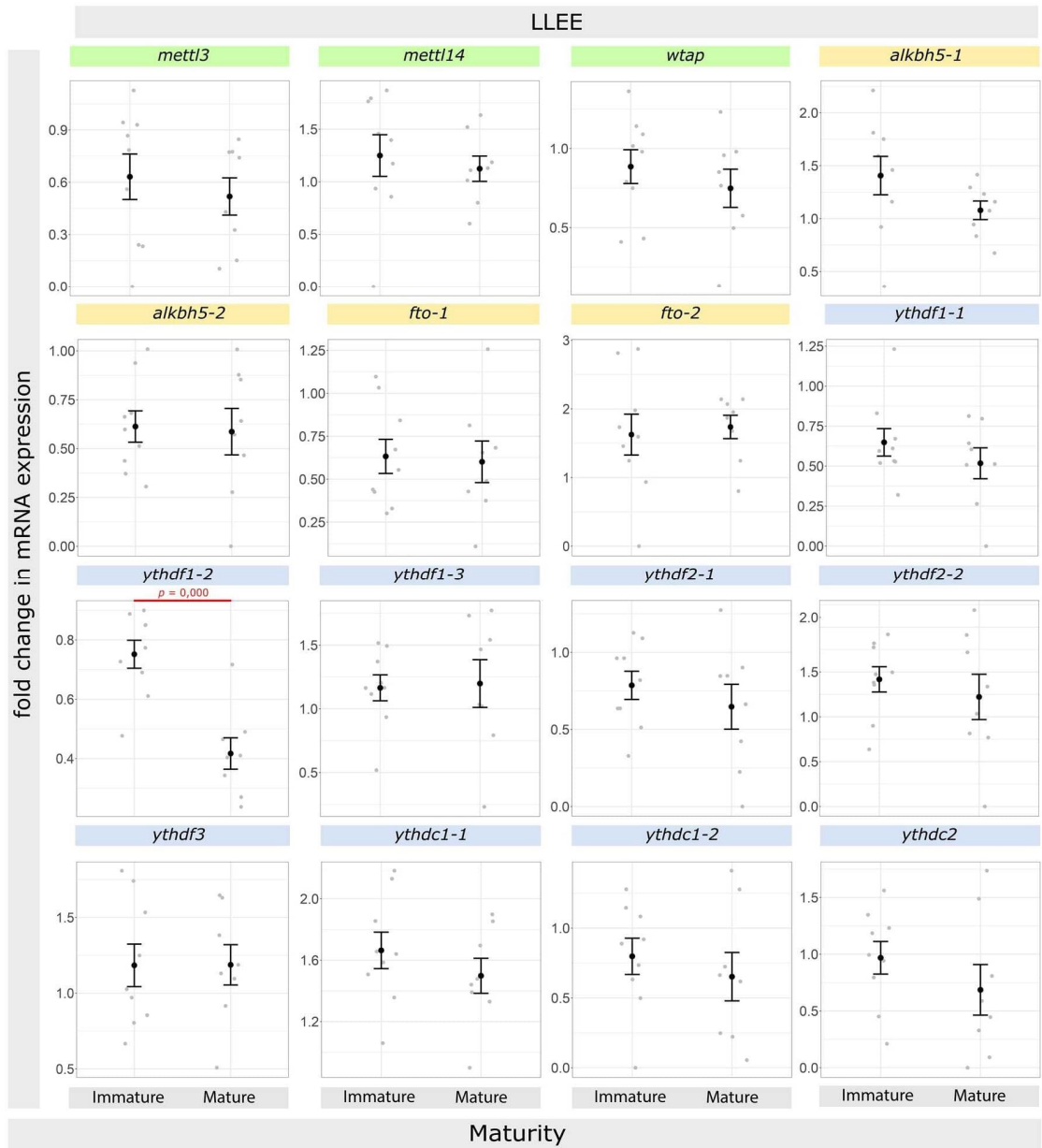

**Figure 4: Comparison of the m⁶A RNA modification regulators mRNA expression level between maturity stages in individuals with LL EE genotypes.** For each gene, values were Log2 Fold-Change of the mRNA expression for each sample (grey dot) and mean ± SEM (black dot and bar). The mRNA level expression differences between the genotypes were analyzed with ANOVA and corrected for multiple comparison using the Benjamini-Hochberg method. p: FDR-adjusted p-values.

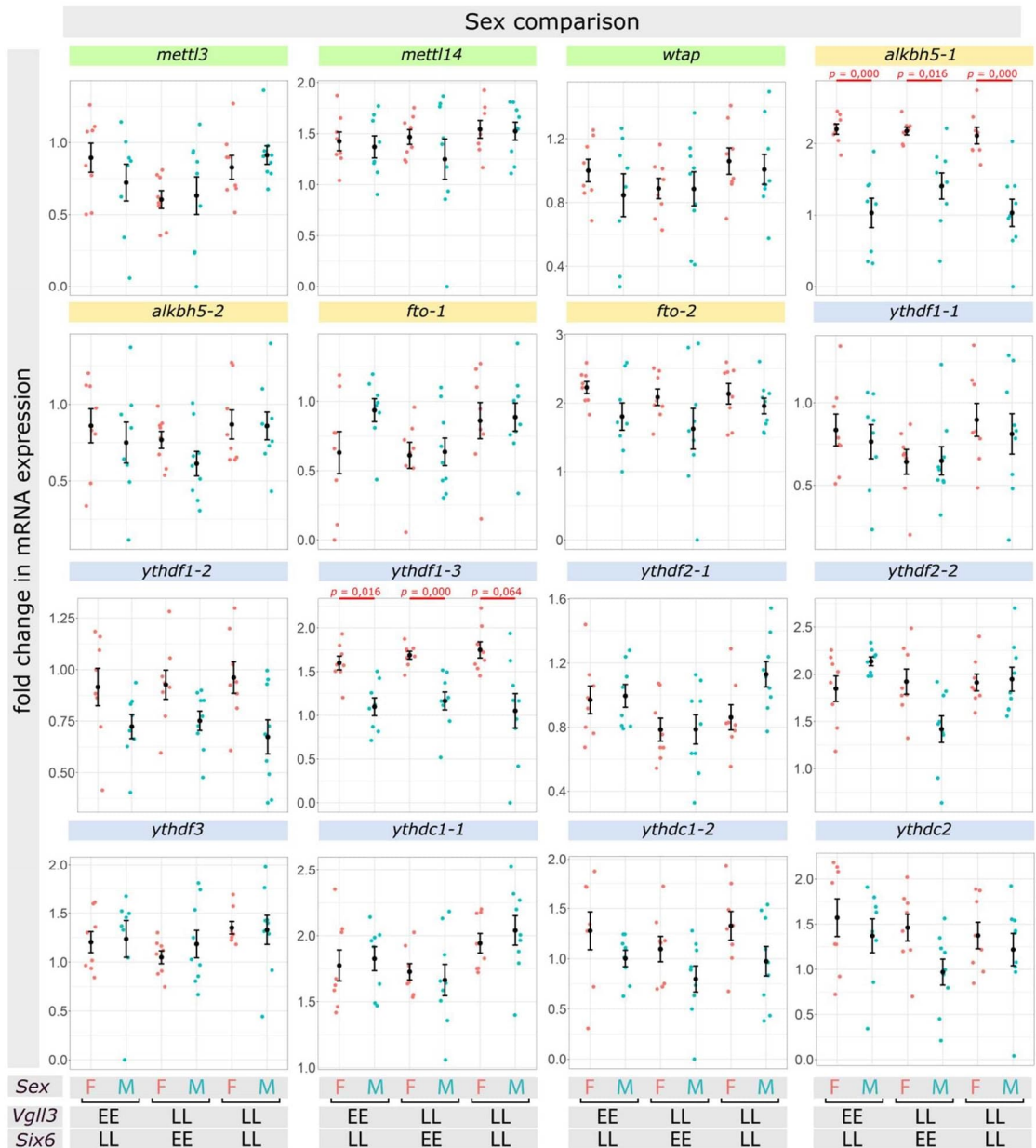

**Figure 5: Comparison of the m⁶A RNA modification regulators mRNA expression level between sexes.** For each gene, values were Log2 Fold-Change of the mRNA expression for each sample (red and blue dots) and mean ± SEM (black dot and bar). The mRNA level expression differences between the genotypes were analyzed with ANOVA and corrected for multiple comparison using the Benjamini-Hochberg method. p: FDR-adjusted p-values.

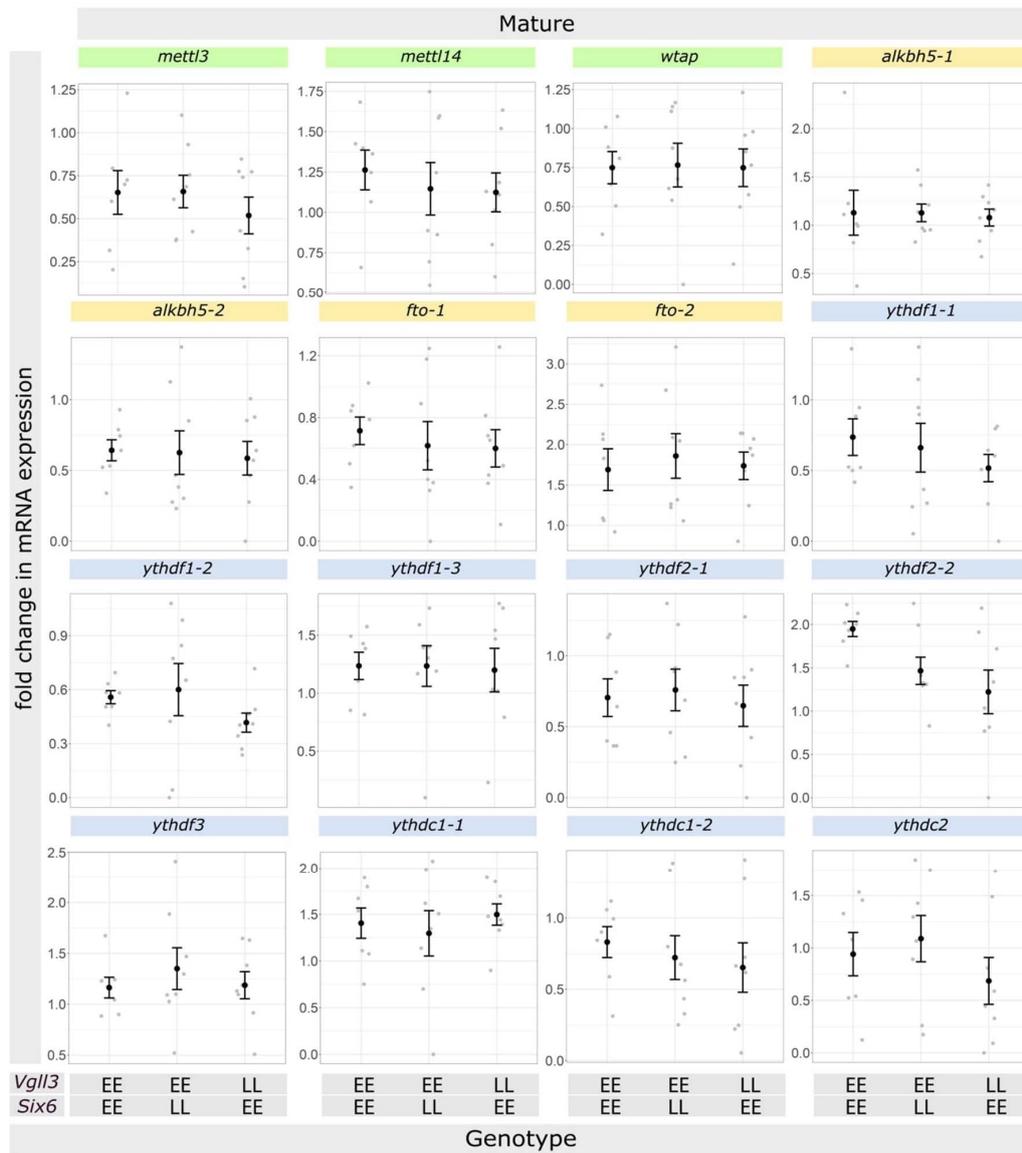

**Supplementary Figure 1: Comparison of the m⁶A RNA modification regulators mRNA expression level between the genotypes in mature males.** For each gene, values were Log2 Fold-Change of the mRNA expression for each sample (grey dot) and mean ± SEM (black dot and bar). The mRNA level expression differences between the genotypes were analyzed with ANOVA and corrected for multiple comparison using the Benjamini-Hochberg method.

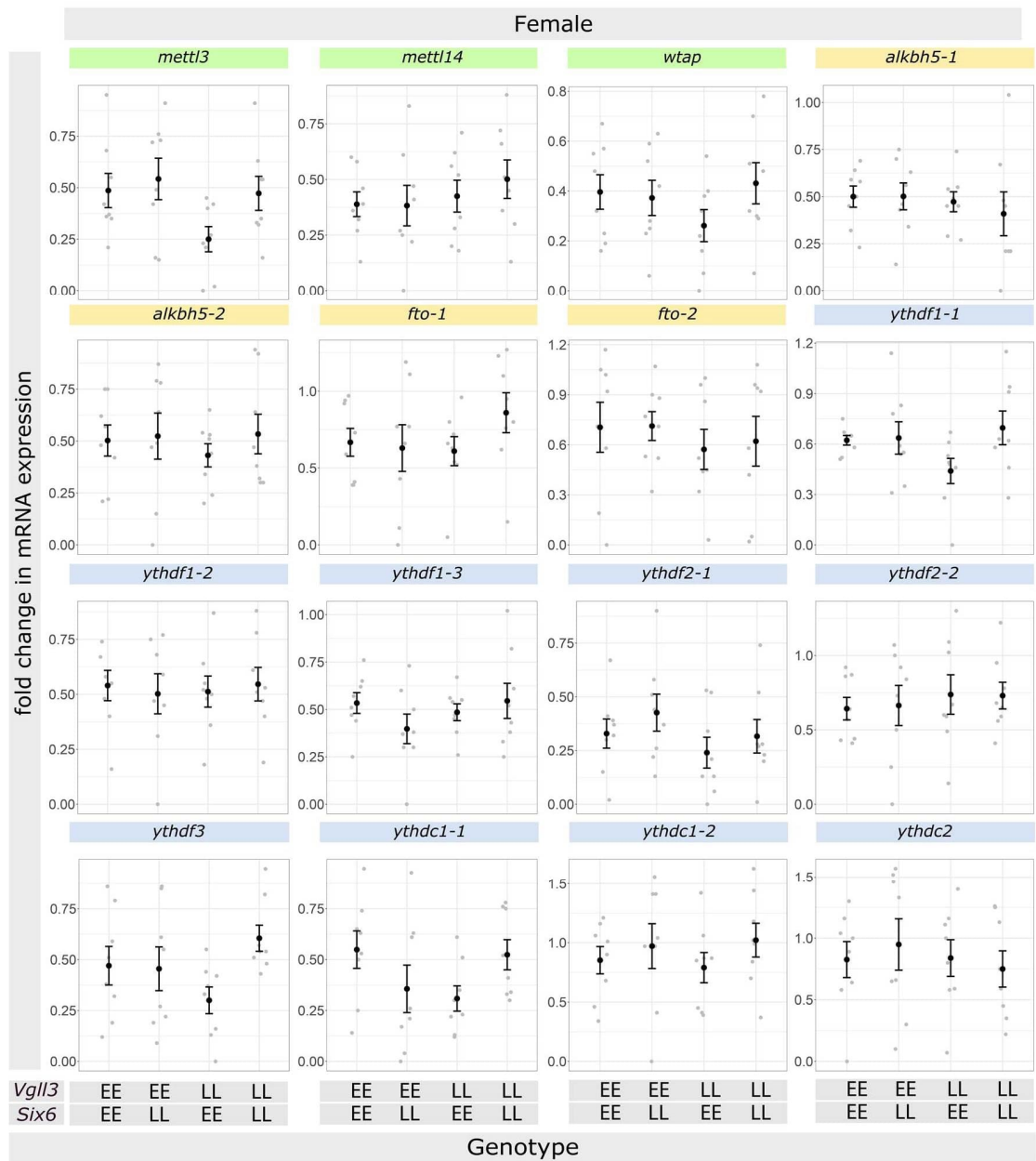

**Supplementary Figure 2: Comparison of the m$^6$A RNA modification regulators mRNA expression level between the genotypes in females.** For each gene, values were Log2 Fold-Change of the mRNA expression for each sample (grey dot) and mean ± SEM (black dot and bar). The mRNA level expression differences between the genotypes were analyzed with ANOVA and corrected for multiple comparison using the Benjamini-Hochberg method.

**Supplementary Data: Expression data and statistical analysis.**

[https://ars.els-cdn.com/content/image/1-s2.0-S0044848623009250-mmc1.xlsx](https://ars.els-cdn.com/content/image/1-s2.0-S0044848623009250-mmc1.xlsx)